\documentclass[12pt]{article} 
\usepackage{graphicx}
\usepackage{cite}
\newcommand\req[1]{Eq.~(\ref{#1})}

\newcommand\rif[1]{Fig.~\ref{#1}}

\newcommand\sr{S_{\rm R}}
\newcommand\tr{T_{\rm R}}

\newcommand\dpar[2]{\frac{\partial #1}{\partial #2}}
\newcommand\rcite[1]{Ref.~\cite{#1}}
\newcommand\kss{K_{S \bar{S}}}
\newcommand\ktt{K_{T \bar{T}}}

\def\simlt{\stackrel{<}{{}_\sim}}

\def\be{\begin{equation}}
\def\ee{\end{equation}}
\def\br{\begin{eqnarray}}
\def\er{\end{eqnarray}}
\def\NPB#1#2#3{{\it Nucl.~Phys.} {\bf{B#1}} (#2) #3}
\def\PLB#1#2#3{{\it Phys.~Lett.} {\bf{B#1}} (#2) #3}
\def\PRD#1#2#3{{\it Phys.~Rev.} {\bf{D#1}} (#2) #3}
\def\PRL#1#2#3{{\it Phys.~Rev.~Lett.} {\bf{#1}} (#2) #3}

\def\MPLA#1#2#3{{\it Mod.~Phys.~Lett.} {\bf{A#1}} (#2) #3}

\def\JHEP#1#2#3{{\it JHEP} {\bf #1} (#2) #3}
\begin{document}
\begin{titlepage}
\vspace{3cm}
\title{Moduli Evolution in Heterotic Scenarios}
\vspace{3cm}
\author{T. Barreiro\thanks{e-mail address: mppg6@pact.cpes.susx.ac.uk.
Address after October $1^{\mathrm{st}}$ 2000:
Centro Multidisciplinar de Astrof\'{\i}sica, 
Instituto Superior T\'ecnico,
Lisboa 1049-001, Portugal.} \\
{\em Centre for Theoretical Physics, University of Sussex}  \\
{\em Falmer, Brighton BN1 9QJ, UK} \\ 
\vspace{0.5cm} \\
B. de Carlos\thanks{e-mail address: Beatriz.de.Carlos@cern.ch. Address after
October $1^{\mathrm{st}}$ 2000: Centre for Theoretical Physics, University of Sussex,
Falmer, Brighton BN1 9QJ, UK.} \\
{\em Theory Division, CERN, CH-1211 Geneva 23, Switzerland} \\
\vspace{0.5cm} \\
N.J. Nunes\thanks{e-mail address: kap13@pact.cpes.susx.ac.uk} \\
{\em Centre for Theoretical Physics, University of Sussex}  \\
{\em Falmer, Brighton BN1 9QJ, UK} }
\date{}
\maketitle
\begin{abstract}
\noindent
We discuss several aspects of the cosmological evolution of moduli 
fields in heterotic string/M-theory scenarios. In particular we study 
the equations of motion of both the dilaton and overall modulus of these 
theories in the presence of an expanding Universe and under different 
assumptions. First we analyse the impact of their couplings to matter 
fields, which turns out to be negligible in the string and M-theory 
scenarios. Then we examine in detail the possibility of scaling in 
M-theory, i.e. how the moduli would evolve naturally to their minima 
instead of rolling past them in the presence of a dominating background. 
In this case we find interesting and positive results, and we compare 
them to the analogous situation in the heterotic string.

\end{abstract}
\thispagestyle{empty}

\vspace{1.5cm} 
\noindent
CERN-TH/2000-302 \\
SUSX-TH/00-014 \\
September 2000

\vskip-22cm
\rightline{CERN-TH/2000-302}
\vskip3in
\end{titlepage}

\newpage

\section{Introduction}

For a number of years now, our knowledge of the string world has greatly 
improved with the discovery of M-theory as the origin of all the known 
perturbative string theories \cite{Horav96p}. In particular, the 
$D=11$ supergravity (SUGRA) limit of this theory (also known as M-theory 
itself) has been extensively studied and, by now, we have quite a deep 
understanding of many aspects of it. 

However the issue of how to stabilize the moduli in these theories, and the 
role they play in cosmology, is still an open one. This is because, in order 
to generate a potential for the moduli, we must rely on non-perturbative 
physics which is not under our control. Still, a few general results can be 
obtained, which encourage further investigation. In particular, supersymmetry 
(SUSY) breaking by gaugino condensation~\cite{Horav96} in the $D=4$ effective 
theory works in a similar way as in the old heterotic string~\cite{Deren85}. 
In fact it is possible to
build many scenarios where multiple gaugino condensates contribute to create
a scalar potential for the moduli $S$ and $T$ that has minima at the 
phenomenologically desired values (${\rm Re} \; S, {\rm Re} \; T \sim O(20)$ 
in $M_{\rm P}$ units)~\cite{Choi98}. Also the fact that the gaugino 
condensates depend 
exponentially on the gauge kinetic functions ($\sim e^{-\alpha_a f_a}$, where
$a$ labels each condensate, $\alpha_a$ are related to the 1-loop beta function
coefficients and $f_a$ are the gauge kinetic functions), with $4 \pi f_a= 
S- \beta_a T$, instead of $f_a=k_a S$ as it was the case in the heterotic string, 
gives rise to the appearance of flat directions in the scalar potential, 
opening up the possibility of inflation along those. This was thoroughly 
studied in Ref.~\cite{Barre00}. 

Despite of these promising results, there is a second issue we want to 
address in this letter, which is the dynamical evolution of the other moduli,
i.e. those which do not have a flat direction and therefore would suffer from 
a `runaway problem', given the exponential nature of their potential. This 
problem, first pointed out by Brustein and Steinhardt~\cite{Brust93} in the
context of the heterotic string, has been studied along the years and several
mechanisms have been proposed in order to alleviate it (see 
Refs~\cite{Horne94,Barre98p,Huey00}). Our goal will be, exploiting the 
similarity between the heterotic string and M-theory scenarios pointed out 
above, to examine this mechanism for dynamically stabilizing moduli in the
context of heterotic M-theory. 

In order to do so, in section~2 we revise the status of the runaway problem in
the heterotic string, with particular emphasis on the recently proposed 
mechanism of Huey et al.~\cite{Huey00}, and its applicability to 
string models. In section~3 we discuss the M-theory setup: first we study
the existence of scaling solutions  in the presence of a dominating background 
and then we consider the above-mentioned couplings of the moduli to matter. 
Finally, we conclude in section~4.

\section{Scaling solutions revisited}

Let us briefly review the issue of scaling solutions in gaugino condensation
models. It is by now a standard result in cosmology
that scalar fields with an exponential scalar potential have
scaling solutions that are global attractors~\cite{Wette88,Ferre97,Copel98}.
In these scaling solutions the field will evolve with the same
equation of state as the background energy density.
As was explained in Ref.~\cite{Barre98p}, the dilaton, $S$, in
the heterotic string theory is a perfect example of such scaling behaviour.
This scalar particle has an exponential type potential which is steep 
enough to make the field roll past its minimum. However, if the energy 
density of the Universe includes some barotropic fluid coupled to the 
dilaton only gravitationally, this field will reach a constant velocity at 
late times, and will
evolve slowly down the exponential slope. This behaviour, however, can only
stabilize the dilaton if it starts its evolution to the left of its minimum.

More recently a new mechanism has been proposed~\cite{Huey00} which could help
stabilizing moduli with exponential potentials both in their strong and weak
coupling regimes. It relies on the existence of couplings of these moduli
(generically denoted by $\phi$, with an exponential type potential
$V_0(\phi)$) to matter fields (which we will call $C$). These can occur either
through the kinetic terms (which, following Ref.~\cite{Huey00}, we parametrise
as $f(\phi) \partial_{\mu} C \partial^{\mu} C$) or through the scalar 
potential ($V_1 \sim g(\phi) C^n$). Assuming the $C$ field to be 
homogeneous, we define its energy density and pressure to be respectively,
\br
\rho_C & = & f \dot{C}^2 + V_1 \label{rodef} \;\; ,\\
p_C & = & f \dot{C}^2 - V_1 \label{pdef} \;\; .
\er
The equations of motion for $\phi$ and $\rho$ are then given by
\br 
& & \ddot{\phi} + 3 H \dot{\phi} - \frac{\partial p_C}{\partial \phi} + 
\frac{\partial V_0}{\partial \phi} = 0 \;\;,
\label{evolp} \\
& & \dot{\rho_{C}} = - \left( 3 H + \frac{\dot{f}}{2 f} \right) (\rho_C + p_C)
 + \frac{\dot{g}}{2 g} (\rho_C - p_C) \;\;,
\label{evolrho}
\er
with $H^2 = \frac{1}{3} (\dot{\phi}^2/2 +V_0 + \rho_C)$.
To be precise, we will consider the case where the $C$ field has coherent 
oscillations about its minimum. In this case, the energy density and 
pressure, averaged over one oscillation, will obey the equation of state 
$p_C = w_C \rho_C$, with $w_C = (n-2)/(n+2)$ from the virial theorem.
We do this in order to have a simple barotropic fluid
as a background. A similar situation could be obtained if we considered
instead the field $C$ to be in thermal equilibrium~\cite{Huey00}.

With these assumptions, the pressure derivative in \req{evolp} becomes
\be
\frac{\partial p_C(\phi,C,\dot{C})}{\partial \phi} =
\left((1+w_C)  \frac{1}{f} \frac{\partial f}{\partial \phi} -  
(1-w_C) \frac{1}{g}
\frac{\partial g}{\partial \phi} \right) \frac{\rho_C}{2} \;\;,
\label{dpcds}
\ee
and \req{evolrho} can be exactly solved to give
\be
\rho_C(\phi, a) = \rho_0 \left( \frac{a}{a_0} \right)^{-3(1+w_C)} 
\left( \frac{f}{f_0} \right)^{-(1+w_C)/2}  
\left( \frac{g}{g_0} \right)^{(1-w_C)/2} \;\;,
\label{rho}
\ee
where $a$ is the scale factor of the Universe and the subscript ${}_{0}$ 
denotes initial values of the corresponding functions. Altogether we see that 
the new term $- \partial p_C/\partial \phi$ can be interpreted as a 
contribution to the total scalar potential of the $\phi$ field. In other 
words, the coupled system will be equivalent to a single field system with
a total potential given by
\be
V_{\rm tot} = V_{0}(\phi) + V_{\rm eff}(\phi) \;\;,
\label{eff1}
\ee
with
\be
V_0(\phi) \sim e^{-\alpha \phi} 
\label{tree}
\ee
and
\be
V_{\rm eff}(\phi) = \rho_C(\phi) \;\;.
\label{eff2}
\ee
Notice that $\rho_C$ is now given by \req{rho}, an explicit function of 
time (or the scale factor, $a$), as opposed to the $\rho_C$ we had 
{\em before} integrating out the $C$ field. They have both the same value 
of course, but their partial derivatives with respect to $\phi$ will be 
different. In particular, it is easy to show that
\be
\dpar{ \rho_C(\phi, a)}{\phi}
  = - \dpar{ p_C(\phi, C, \dot{C})}{\phi} \;\;,
\label{devrel}
\ee
the result we have used to obtain \req{eff2}. This effective potential
is represented in Fig.~1a, where we show the shape of $V_{\rm tot}$ as
a function of $\phi$, for several values of $N \equiv \ln(a)$ with 
$w_C = 1/3$, for the particular case of 
$f(\phi) = g(\phi) = 1/\phi$ considered in~\rcite{Huey00}. Here $V_0$
is given by a two condensate model analogous to Eq.~(4) of Ref.~\cite{Huey00}.
We can see that, at earlier times (i.e. for $N$ small), the potential has a 
growing slope to the right of the minimum, whose position changes with $N$. 
At late enough times (i.e. for $N$ large) this effective contribution fades 
away and we recover the original exponential potential $V_0(\phi)$. As we 
see in Fig.~1b, even for initial conditions in the very weak coupling 
regime (characterised by large values of $\phi$), the field rolls towards 
its minimum\footnote{It can be argued that, in the weak coupling regime, 
it would be inconsistent to consider the $C$ field in thermal equilibrium 
with the background. However this is beyond the scope of this paper which
is to show that in a string motivated scenario the effect of matter fields is
negligible in any case.}.

\begin{figure}
\centerline{  
\includegraphics[width=8cm]{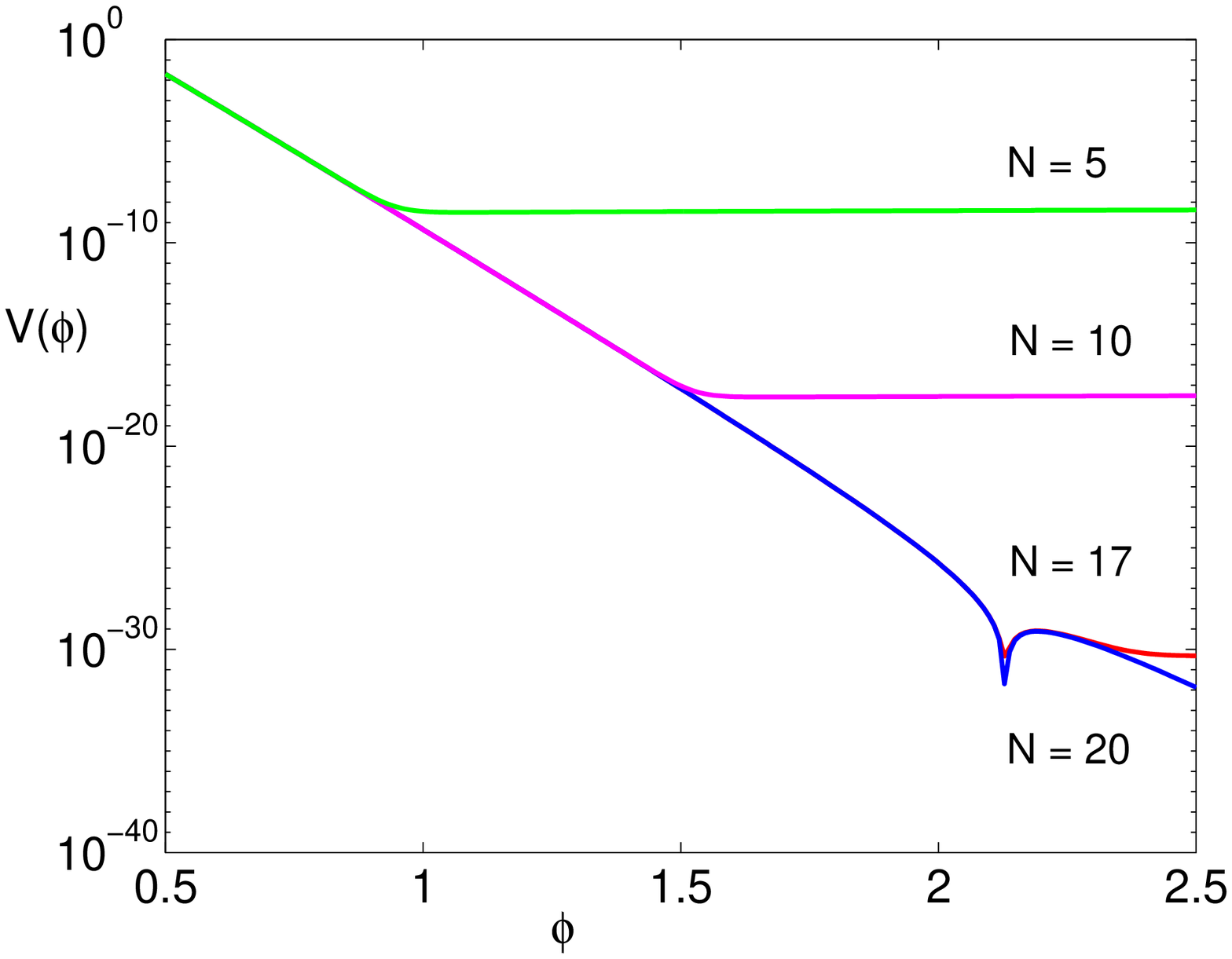}
\includegraphics[width=8cm]{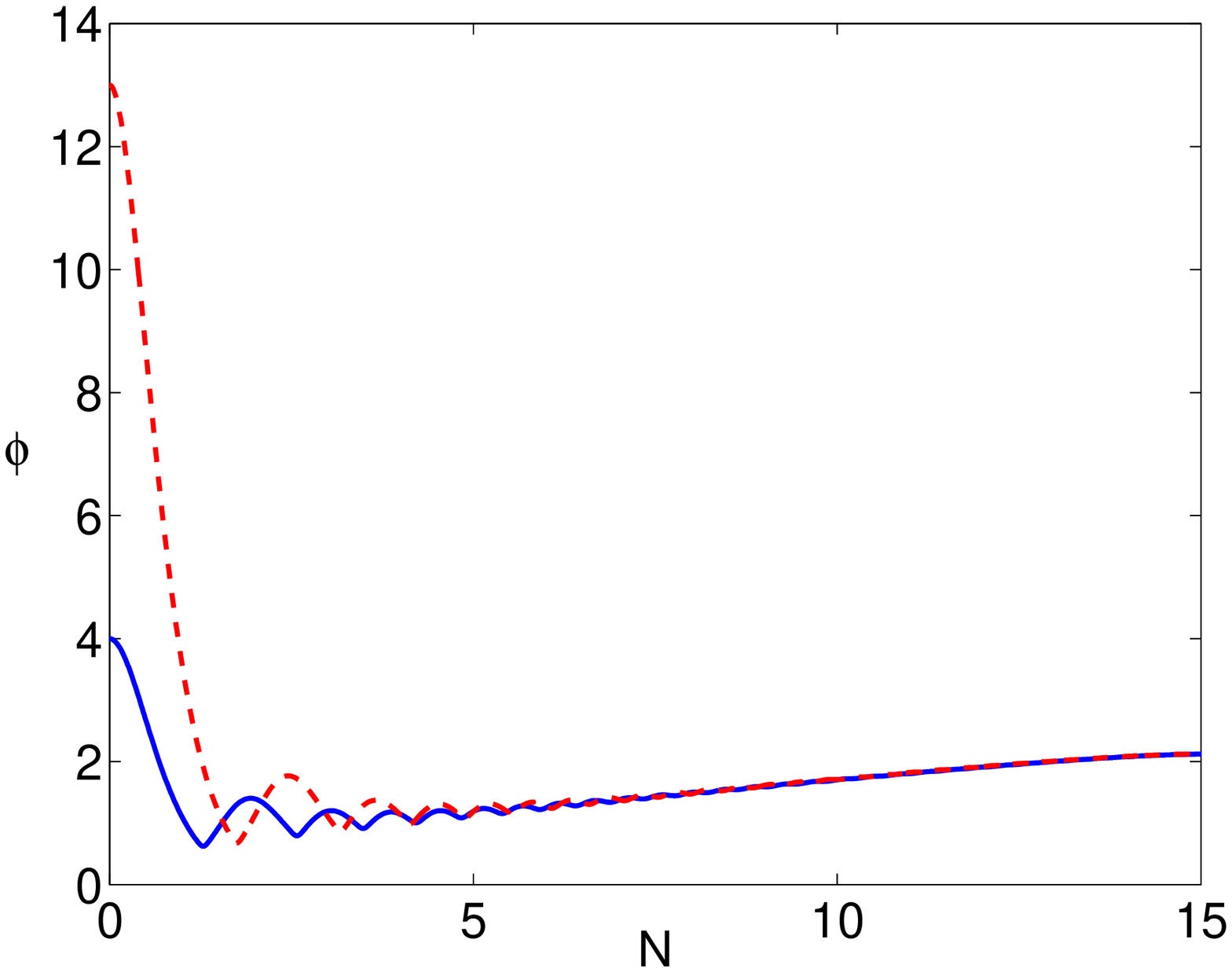}
}
\caption{}
{\footnotesize \noindent \textbf{\sffamily (a)} Plot of the total scalar 
potential, given by Eq.~(\ref{eff1}), as a function of the dilaton $\phi$ 
(in $M_P$ units) in a two condensate model with groups ${\rm SU}(6)_2 \times
{\rm SU}(7)_8$ and $T_{\rm R} = 1.23$ (see footnote~2). The different 
curves correspond to different times -or number of $e$-folds $N$- in the 
evolution of the system. 
\textbf{\sffamily (b)} Evolution of the dilaton ($\phi$), as in 
Eq.~(\ref{evolS}), as a function of $N \equiv \ln(a)$, for the same example 
as in Fig.~1a and initial conditions given by $\phi_0=13$ (dashed), 
$\phi_0=4$ (solid).}
\label{fig1}
\end{figure}

Despite of how promising this mechanism looks as a way of stabilizing 
moduli, it is very sensitive to the 
particular forms of the functions $f$ and $g$ (recall $f=g=1/\phi$ in
the above example). Therefore 
it is a natural step to proceed with the study of string inspired models and, 
in particular, we shall devote some time to consider the dilaton of the 
heterotic string and its couplings. 

The first thing one must take into account is that, in the context of
the heterotic string, the dilaton $S$ is not a canonically normalised
field, i.e. its kinetic term is given by $K_{S \bar{S}} \partial_{\mu} S 
\, \partial^{\mu} {\bar S}$, with $K_{S \bar{S}} = \partial^2 K/\partial S 
\partial \bar{S}$ and the tree-level K\"ahler potential given by 
$K=-\ln(S+\bar{S})$. Therefore its equation of motion reads
\be
\ddot{S} + 3 H \dot{S} + \frac{K_{S \bar{S} \bar{S}}}{K_{S\bar{S}}} 
\dot{S}^2 - \frac{1}{K_{S \bar{S}}} \frac{\partial p_C}{\partial {\bar S}} + 
\frac{1}{K_{S \bar{S}}} \frac{\partial V}{\partial {\bar S}} = 0 \;\;,
\label{evolS}
\ee
where $K_{S \bar{S} \bar{S}} = \partial K_{S\bar{S}} / \partial \bar{S}$ and
$H^2 = \frac{1}{3} (K_{S\bar{S}} \dot{S}^2 + K_{C\bar{C}} \dot{C}^2 + 
V(S,C))$. It must be said, however, that the term proportional to 
$\dot{S}^2$ is negligible in the region of values of $S$ studied here, and 
therefore this equation is very similar to the simpler
Eq.~(\ref{evolp}) described previously. The couplings to matter,
contained in 
$\partial p_C/ \partial S$, are determined by the structure of the 
N=1 SUGRA Lagrangian that describes this string-inspired
model\footnote{Note that, throughout this section in which we focus
on the evolution of the dilaton, we are assuming that the $T$ modulus, 
whose vacuum expectation value (VEV) represents the overall size of the 
compactified space, is fixed 
to its value at the minimum. We will include the $T$ modulus explicitly
when considering heterotic M-theory scenarios in the next section.}.
The kinetic term for $C$ reads~\cite{Lust92} 
\be
K_{C \bar{C}} \partial_{\mu} C \partial^{\mu} \bar{C} = \frac{3}{2 \tr} 
\left( 1+ \frac{\epsilon}{6 \sr} \right) \partial_{\mu} C \partial^{\mu} 
\bar{C} \;\;,
\label{kinc}
\ee
where $\epsilon = \delta_{GS}/4 \pi^2$ is the so-called Green--Schwarz 
coefficient, which is generally small, and $\sr \equiv {\rm Re} S$, 
$\tr \equiv {\rm Re} T$. Therefore in the parametrisation introduced 
before \req{rodef}, $f(\sr)$ is given by
\be
f(\sr) = \frac{3}{2 \tr} \left( 1+ \frac{\epsilon}{6 \sr} \right) \;.
\ee
Finally, the form of the scalar potential is given by
\be
V = e^K \left\{ \sum_{i,j} (W_i + K_i W) {(K_i^j)}^{-1} (\bar{W}^j +  
K^j \bar{W}) - 3 |W|^2 \right\} \;\;,
\label{tpotential}
\ee
where sub(super)scripts denote  
derivatives of $W$ and $K$ with respect to $S$ ($\bar{S}$)
and $C$ ($\bar{C}$), and  $W \sim \sum_a e^{-\alpha_a S} + C^3$
is the superpotential coming from multiple gaugino condensation.
The couplings 
between the two fields $S$ and $C$ arise from the term 
$e^K |K_C W+W_C|^2/K_{C\bar{C}}$. For $C$ small,
$W_C \sim C^2$ will dominate and we can identify 
\be
g(\sr) = e^K/K_{C\bar{C}} = \frac{1}{24 \tr^2 \sr} \;.
\ee
Therefore this coupling is of the form $1/\sr$.

We can now solve Eq.~(\ref{evolS}), and the result is that $\sr$ has the 
same runaway behaviour that it would have had in the absence of the 
couplings to matter. Let us briefly sketch why the mechanism of 
Ref.~\cite{Huey00} does not work here: consider the general form for the 
couplings $f=a+b/\sr$, $g=c+d/\sr$, with $a$, $b$, $c$, $d$ real and 
positive. A slope to the right of the minimum, such as those in Fig.~1a,
will appear if, in Eq.~(\ref{evolS}), we require $- \partial p_C/ 
\partial S > 0$ to balance $\partial V/\partial S$ which is invariably 
negative (recall that $V \sim e^{-\alpha S}$). From Eq.~(\ref{dpcds})
is it easy to translate this condition, for {\em positive} $w_C$,
into either of these two
\br
i) \; ad(1-w_C)-bc(1+w_C) & \leq & 0 \nonumber \\
\label{conds} \\
ii) \; ad(1-w_C)-bc(1+w_C) & > & 0 \;\; {\rm and} \;\; 
\sr < \frac{2bdw_C}{ad(1-w_C)-bc(1+w_C)} \;. 
\nonumber 
\er
In Ref.~\cite{Huey00}, indeed $w_C=1/3$ is positive and $f$ and
$g$ are such that $a=c=0$, that is this choice 
fulfils condition {\em i)} and their $\phi$ field is always trapped at 
the minimum. In the heterotic string, however, $c=0$ and $a \neq 0$, so 
we are in case {\em ii)}, which implies that the dilaton will be trapped
only for initial conditions $\sr \simlt b/a = \epsilon/6 \ll 1$, for any
values of $w_C \neq 1$. This 
corresponds to the strong coupling regime and covers a very small
fraction of the parameter space. For completeness we can also consider
the case in which $-1 \leq w_C \leq 0$, and now we will have
only one condition for $- \partial p_C/ \partial S > 0$, which is
\be
 ad(1-w_C)-bc(1+w_C)  <  0 \;\; {\rm and} \;\; 
\sr > \frac{2bdw_C}{ad(1-w_C)-bc(1+w_C)} \;. 
\label{condp}    
\ee    
It is easy to see that, again, for the heterotic string where $c=0$,
this condition can never be fulfilled.     
     
Therefore to end this section we should conclude that,
concerning the probability of 
stabilizing the moduli in the Early Universe, the estimates of Horne 
and Moore \cite{Horne94} still represent the most optimistic
situation. They argued that the motion of the moduli follows
a chaotic trajectory and, out of the finite volume of possible initial 
conditions, 12-14\% of it corresponds to cases where the dilaton will end 
up at its minimum. However, in Ref.~\cite{Banks95} it was pointed out how, 
in their scenario, the inhomogeneous modes would soon come to dominate the 
energy density of moduli fields. Finally, if we consider the presence of a 
dominating background, then the estimates of Ref.~\cite{Barre98p}
remain unaltered in the presence of above-mentioned couplings to 
matter fields.

\section{Dynamics of the moduli in M-theory}

In this section we will try to apply the same ideas, namely the evolution
of moduli fields in the presence of some background matter, to study the 
evolution of the real parts of the $S$ and $T$ moduli in heterotic M-theory.
From a phenomenological point of view, one of the main differences between
the two models is that the gauge kinetic functions in the hidden wall,
$f_a$  (whose VEV determine the gauge coupling constants) are now a linear 
combination of the string dilaton and modulus fields, i.e 
$f_a=(S - \beta_a T)/(4 \pi)$ where $\beta_a$ are constants that depend
on the details of the model and are usually of order one \cite{Choi97} (in 
particular, in all the examples shown in this paper we have fixed $\beta_a=
\beta=1/2$).
Assuming once again that gaugino condensation is the source of SUSY breaking,
the scalar potential will have an exponential like profile, as in the string 
case, along the direction defined by $\Phi_- \equiv S - \beta T$. Along its 
orthogonal direction, $\Phi_+ \equiv \beta S + T$, the potential will be 
almost flat, the only dependence upon this variable coming from the K\"ahler 
potential $K$. We will consider potentials where a minimum is generated along 
the $\Phi_-$ direction with two gaugino condensates. This defines a 
`valley' in the $\Phi_+$ direction with an inverse power law profile, and 
the global minimum is obtained with a third condensate, or a non-perturbative 
correction to the K\"ahler potential \cite{Choi98,Barre00}.

\begin{figure}
\centerline{
\includegraphics[width=10cm]{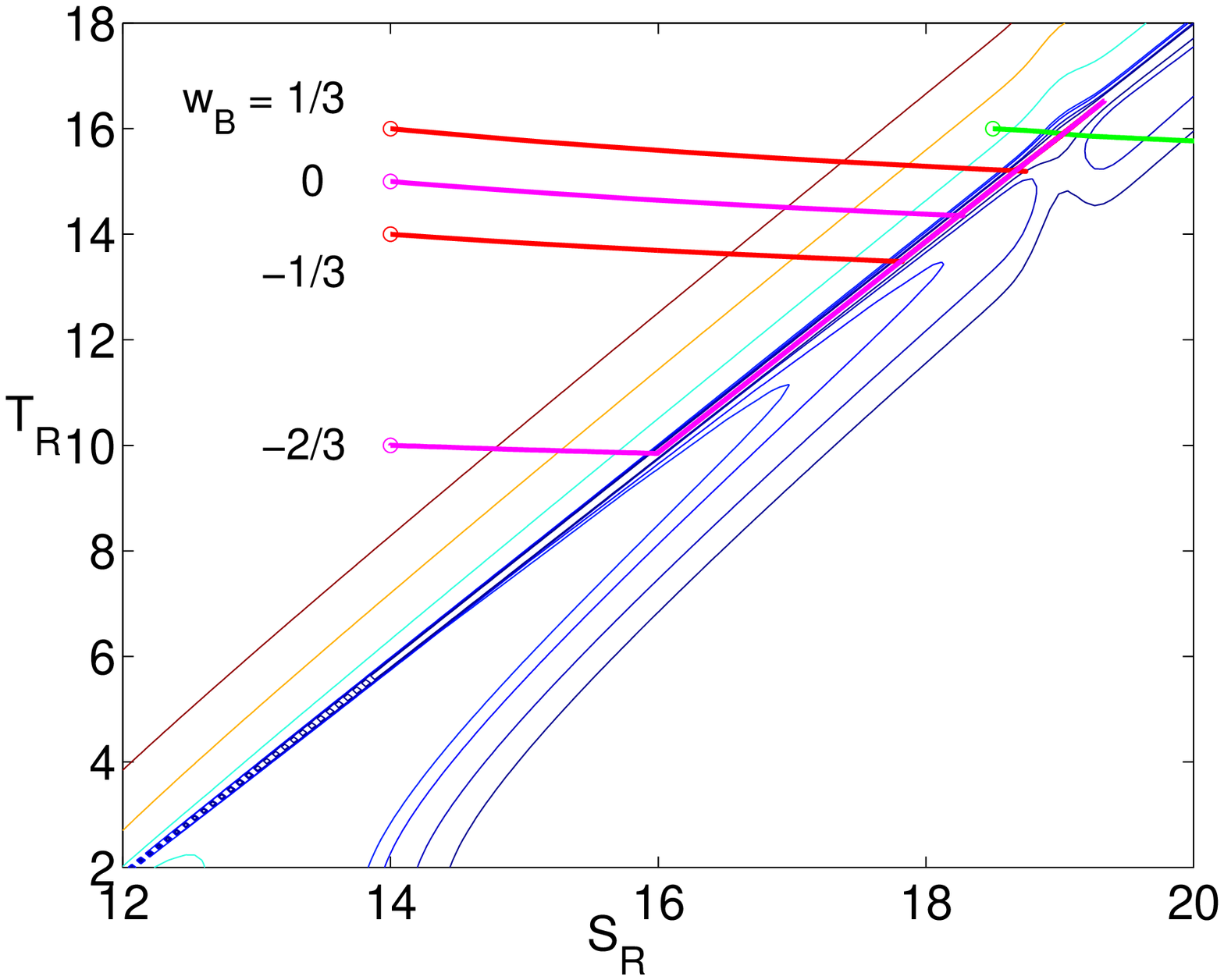}
}
\caption{}
{\footnotesize \noindent \textbf Contour plot of the scalar potential
$V$ in terms of $\sr$, $\tr$. The thick lines represent the trajectories
of the fields in the presence of the corresponding backgrounds, in
particular the upper right line corresponds to no background present.
The parameters used for $K_{\rm np}$ are $B = 120$ and $D = 0.13$. One can
see that when a background fluid is present, the range of initial conditions
allowing stabilization is highly improved. 
}
\label{conplot}
\end{figure}

Let us start with the simplest case, when the $S$ and $T$ fields have no 
direct coupling to the background fluid. As we have just said we have 
essentially two major directions in our scalar potential, an exponential 
slope for the $\Phi_-$ field and an inverse power-law for the $\Phi_+$ field. 
In the exponential part of the potential we expect the field $\Phi_-$ to 
dominate the evolution, behaving in a similar way to the dilaton in the 
heterotic string case \cite{Barre98p}.
However, in the valley defined by $\Phi_-=\Phi_-^{\rm min}$, we expect the 
evolution to be driven by the $\Phi_+$ field along the valley. Therefore the 
stabilization of the fields will rely on a two-step process, first the 
evolution into the valley, and then the stabilization along the valley
to the minimum (see Fig.~\ref{conplot}). 

We will analyse the behaviour of the fields in this model using a
numerical simulation with a specific example. The equations of motion of the 
two fields, with no direct couplings to the background, is given by
\br
\ddot{S} + 3 H \dot{S} + \frac{K_{S \bar{S} \bar{S}}}{K_{S \bar{S}}} \dot{S}^2
  + \frac{1}{K_{S \bar{S}}} \frac{\partial V}{\partial {\bar S}} & = & 0 \;\;
\label{evolS2} \\
\ddot{T} + 3 H \dot{T} + \frac{K_{T \bar{T} \bar{T}}}{K_{T\bar{T}}} \dot{T}^2
 + \frac{1}{K_{T \bar{T}}} \frac{\partial V}{\partial {\bar T}} & = & 0 \;\;,
\label{evolT2}
\er
with the Hubble constant now being given by
\be
H^2 = \frac{1}{3}( K_{S \bar{S}} \dot{S} \dot{\bar{S}} 
     + K_{T \bar{T}} \dot{T} \dot{\bar{T}}
     + V + \rho_{\rm B} ) \;\;,
\label{hub}
\ee     
where $\rho_{\rm B}$ is the background energy density. We evolve both fields 
simultaneously, since the flat directions in the scalar potential do not 
coincide with the ones in the K\"ahler potential.
As detailed in \rcite{Barre00}, the scalar potential is explicitly given by
\be
V = e^K \left\{ \frac{1}{\kss}   |W_S + K_S W|^2 + \frac{1}{\ktt} 
|W_T + K_T W|^2 - 3 |W|^2 \right\} \;\;,
\label{potential}
\ee
where $W$, the superpotential for multiple gaugino condensation, reads as
\be
W = \sum_{a} C_a e^{-\alpha_a \Phi_-} \;,
\ee
with $C_a$ and $\alpha_a$ being constants related to the one-loop
beta-function coefficients of the corresponding condensing groups. We 
will consider for the numerical examples the case of two condensates, 
${\rm SU}(3) \times {\rm SU}(4)$, with 8 pairs of matter fields 
transforming as $(4, \bar{4})$. The K\"ahler potential is given by
\be
K(\sr,\tr) = -\log(S + \bar{S}) - 3 \log(T + \bar{T})
+ K_{\rm np}(\sr) \;\;,
\ee
where the first two terms are the tree level result and $K_{\rm np}$ accounts 
for non-perturbative corrections. We will use for the numerical examples 
the ansatz for $K_{\rm np}$ introduced in \rcite{Barre98a},
\be
K_{\rm np} = \frac{D}{B \sqrt{S_{\rm R}}} \log \left(
1 + e^{-B ( \sqrt{S_{\rm R}} - \sqrt{S_0})} \right) \;\;
\ee
where we fix $S_0 = 19$, and for each particular value of $B$ we choose 
$D$ so that the cosmological constant is equal to zero.

Let us then start analysing the evolution of the fields outside the valley
in the scalar potential. In \rif{eosfi} we show the numerical evolution of
the fields' equation of state, $w_{\Phi}$, as a function of $\Phi_-$ before 
the fields reach the valley at $\Phi_-^{\rm min} = 11.07$. A set of 
different initial conditions and different types of background fluid were
considered.
     
\begin{figure}
\centerline{
\includegraphics[width=10cm]{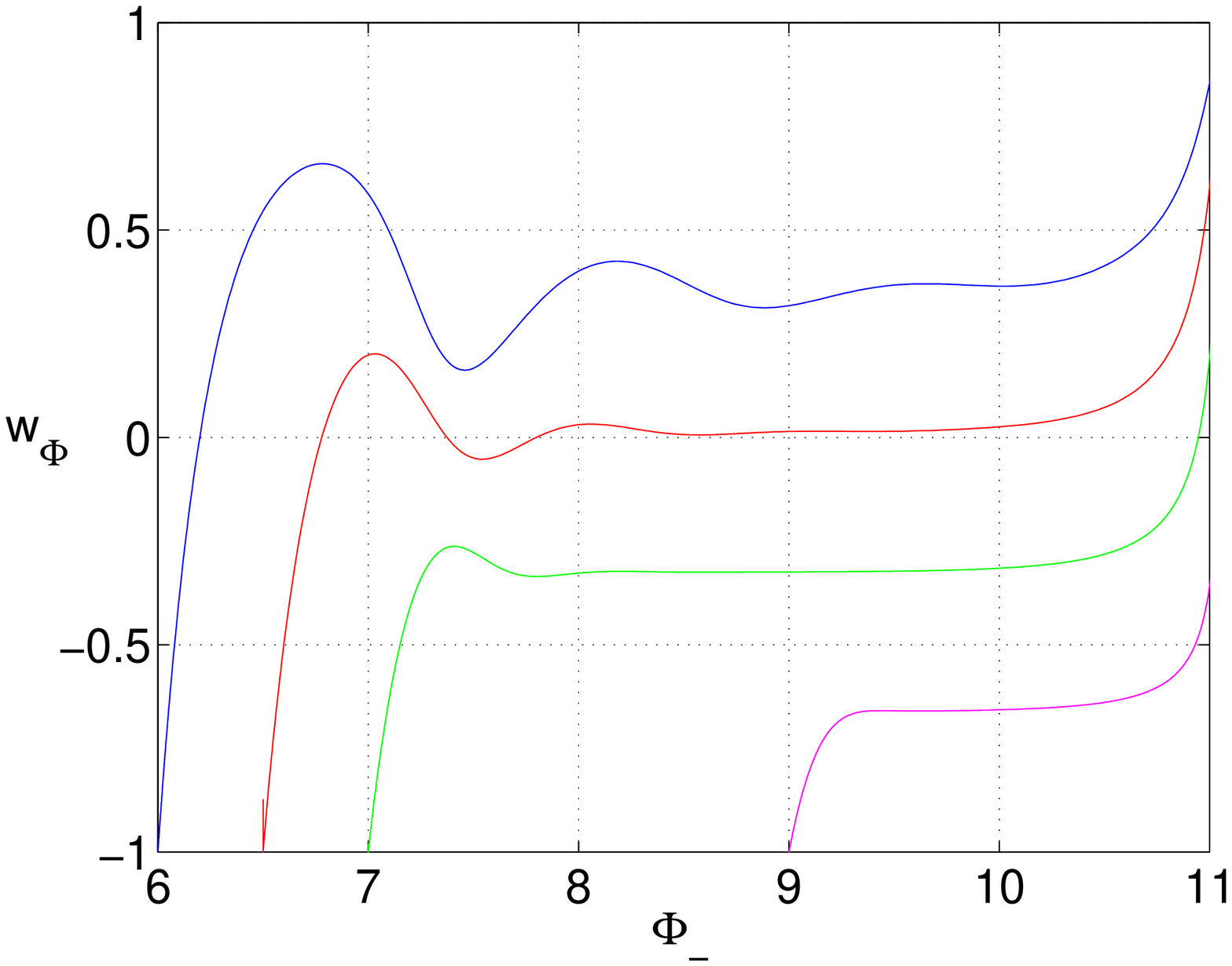}
}
\caption{}
{\footnotesize \noindent \textbf
Equation of state of the system of fields $\Phi_-$, $\Phi_+$ as a
function of $\Phi_-$ before reaching the minimum at $\Phi_-^{\rm min} 
\approx 11.07$. From top to bottom the different curves correspond
to the following choice of background ($w_B$): 1/3, 0, -1/3, -2/3. 
}
\label{eosfi}
\end{figure}


It is easy to see from this plot that the behaviour of the fields
outside the valley is very similar to the case of the dilaton
in the heterotic string case. That is, they will quickly reach a
scaling solution in which their energy density scales with the 
background with roughly the same equation of state (i.e. $w_{\Phi} \sim
w_B \equiv p_B/\rho_B$).
As in the string case, this will be enough to slow down the evolution 
of the fields and allow them to settle in the valley for a wide range 
of initial conditions. However, note in \rif{conplot} that the evolution 
of the fields is not orthogonal to the direction of the valley.
In other words, $\Phi_-$ and $\Phi_+$ scale simultaneously.
The direction of the evolution is determined from the scalar potential
gradient and the K\"ahler potentials of the two fields,
and it is not possible to estimate it analytically.

The main differences with the heterotic string case arise once
the fields have settled in the valley. In order to be stabilized,
they still have to evolve inside the valley towards the minimum.
The valley corresponds approximately to a constant $\Phi_-$, which
we denote as $\Phi^{\rm min}_-$, so that the evolution at this stage 
will be dominated by the $\Phi_+$ field. The effective scalar
potential for the $\Phi_+$ field is power-law like. To be precise, the 
dominating term is the $e^{K}$ factor in \req{potential}, that is
\be
V \approx \frac{V_0}{(\beta \Phi_+ + \Phi_-^{\rm min}) (\Phi_+ - \beta 
\Phi_-^{\rm min})^3} \;\;,
\label{effpot}
\ee
where $V_0$ is a constant that can be estimated analytically (exactly for
the D=0 case) and
$\Phi_-^{\rm min}$ is understood to be held constant.
In this approximation, where $\Phi_-$ is constant, one
can also write an effective kinetic term for the $\Phi_+$ field given in 
terms of the kinetic terms for the $S$ and $T$ fields. Namely, since 
$S = (\beta \Phi_+ + \Phi_-)/(1 + \beta^2)$ and
$T = (\Phi_+ - \beta \Phi_-)/(1 + \beta^2)$, we have that
\be
K_{S \bar{S}} \dot{S}^2 + K_{T \bar{T}} \dot{T}^2 \approx
\frac{ \beta^2 K_{S \bar{S}} + K_{T \bar{T}} }{(1 + \beta^2)^2}
\; {\dot{\Phi}_+}^2 \;\;.
\label{fipkin}
\ee
The final result is a field which has a power-law type potential and
a kinetic term that is inversely proportional to $\Phi_+^2$. If we transform
it into a real field, $\sigma$, with canonical kinetic terms,
$\frac{1}{2} \dot{\sigma}^2$, it is easy to check that $\sigma = 
\sqrt{2}\;\log{\Phi_+}$ and therefore its effective scalar potential will 
be an exponential.
Hence, we expect the behaviour at late times to be the usual scaling solution
associated with exponential potentials. To be more precise, for large values
of the field the effective potential, \req{effpot}, goes as $1/\Phi_+^4$, and
therefore the effective scalar potential for $\sigma$ is $V(\sigma) = 
e^{-2 \sqrt{2} \sigma}$. Without a minimum, that is for $K_{\rm np} =0$, 
     this is
exactly the scaling solution one obtains numerically for late times.

This is a nice result in itself, showing that the fields tend to join 
scaling solutions (albeit ones of different type) when in the presence
of a background, both in the valley and outside it. However, the time
the fields take to reach the scaling solution in the valley is
considerably longer. This means that they will usually reach the
minimum of the potential before acquiring a scaling behaviour. This in 
itself does not stop the fields from being trapped at the minimum, since 
the background is already slowing them down considerably, but it 
does mean that an analytical approach is much more difficult, since the 
field reaches the minimum at an intermediate regime, between prescaling 
and scaling.

We can have an idea of the range of initial conditions that stabilize the 
fields by looking at Table~1. First of all, if the value 
of $\rho_B^0$ is too small, the fields will begin to evolve as if
there was no background (that is, they will not stay in the minimum
unless they start really close to it). Once the background becomes 
important, the actual value of $\rho_B^0$ does not really affect the 
evolution, since the fields will just stay frozen until the energy 
densities of the fields and background become comparable. Therefore we 
chose to fix $\rho_B^0=10^{20}$ in the examples
given in Table~1. The value of $w_B$, on the other hand, affects 
drastically the results. The lower it is, the easier it is for the fields to 
be stabilized. This is not too surprising, since a smaller $w_B$ corresponds 
to a universe where the potential terms dominate over the kinetic terms (and 
recall that the fields will mimic the background once they reach their 
scaling solutions). Moreover, different values of the $(B,D)$ parameters
in the non-perturbative K\"ahler will give slightly different answers. In 
short, the smaller the $B$ parameter, the less fine tuned is the minimum in 
the scalar potential, and the easier it is to stabilize the fields.
Finally, one can check that the further the fields
are from the minimum
when  they hit the valley, the more difficult the stabilization becomes.  
The full evolution for specific examples where stabilization is achieved
can be seen in \rif{conplot}.

\begin{table}[h]
\center{
\begin{tabular}{|c|c|c|c|c|c|}
\hline
{\bf $K_{\rm np}$} {$(B, D)$} & {\bf $\tr^0$} & {\bf $w_B$}
 & {\bf Stability} \\ \hline
(120, 0.13) & 16 & 1/3 & $\surd$ \\
(120, 0.13) & 14 & 0 & $\times$ \\
(120, 0.13) & 14 & -1/3 & $\surd$ \\
(120, 0.13) & 10 & -1/3 & $\times$ \\
(120, 0.13) & 10 & -2/3 & $\surd$ \\
(120, 0.13) & 6 & -2/3 &  $\times$\\
(50, 0.33) & 6 & -2/3 & $\surd$ \\
(50, 0.33) & 6 & -1/3 &  $\times$\\
(10, 2.18) & 6 & -1/3 & $\surd$ \\
(10, 2.18) & 6 & 0 &  $\times$\\
\hline
\end{tabular}
}
\caption{Examples of points in parameter space for which 
the fields are -or not- stabilized at the minimum. In all of them 
$\rho_B^0 = 10^{20}$ and $\sr^0 = 14$.}
\end{table}

Let us finally turn to the case where the dilaton and modulus fields have
direct couplings to the background fields. As for the string case considered
in Section~2, we will again  model the background as a scalar field $C$ with
a coupling $f(\sr,\tr)$ in the kinetic terms, and $g(\sr,\tr)$ in the scalar 
potential. Its energy density and pressure is again given by 
Eqs~(\ref{rodef},\ref{pdef}) and the dilaton and modulus equations of motion 
by \req{evolS2} and \req{evolT2} respectively. 
The couplings are now functions of both fields \cite{Choi97}, namely
\br
f(\sr,\tr) & = & \frac{3}{2 \tr} + \frac{\beta}{2 \sr} \;\; , \label{mfdef} \\
g(\sr,\tr) & = & \frac{1}{4 \tr^2 (3 \sr + \beta \tr)} \;\; . \label{mgdef}
\er
The situation here is similar to the heterotic string case if slightly more 
involved. Again, $\partial V/\partial S$ is always negative (except near 
the valley where the condensates cancel each other), and the $S$ field tends 
to go to infinity. If the interaction term in \req{evolS} is to cancel this 
natural tendency, then we require $\partial p_C/\partial S < 0$. It is easy 
to check that, for positive $S$ and $T$, this is equivalent to
$S < 2 \beta/ 3 T$. This only happens for $S$ smaller than its  minimum 
value, so if the initial conditions are larger than this, the field will 
always run away to infinity. For the $T$ field, however, things do change. 
The derivative term $\partial V/ \partial T$ is always positive (again, 
except near the valley), and therefore the field will tend to go to zero (it 
will never become negative since its kinetic term diverges at zero). It is 
possible to check that $\partial p_C/ \partial T$ has the same sign as $p_C$ 
and therefore, for positive background pressure, the coupling term 
balances the potential term in the equation of motion for $T$. Indeed,
we have checked numerically that the $T$ field will tend to go to a 
non-zero constant in the presence of these coupling terms. This can
also be understood analytically 
if one looks for the asymptotic solutions in this regime. 

Finally let us add that this analysis was performed in the absence
of non-perturbative corrections to the K\"ahler potential, 
$K_{\rm np}(\sr)$ (notice the form of Eq.~(\ref{mgdef})). In the
presence of such corrections, that would only affect the behaviour of the
$S$ field, we would have to replace 
Eq.~(\ref{mgdef}) by $e^{K_{\rm np}} g(\sr,\tr)$, and its derivative 
with respect to $\sr$ would be now $e^{K_{\rm np}} (\partial g/\partial \sr
+ g \partial K_{\rm np} / \partial \sr )$. Given that
$ \partial K_{\rm np} / \partial \sr$ is also negative, this would not
alter the conclusions we just reached about the evolution of $S$.
     
In short, adding the interaction terms only improves the situation for
the $T$ field. In practice, this means that the fields still have to
start to the left of the valley in the scalar potential if they are to 
become trapped at the minimum. We did not find any improvement in the 
allowed regions of parameter space presented in Table~1 when we
included couplings with the matter fields.

\section{Conclusions}

In this letter we have addressed the dynamical evolution of moduli fields
in several string/M-theory scenarios. In all of them the dynamics are
provided by assuming gaugino condensation in the hidden sector of the theory 
as the source of SUSY breaking and, therefore, of a non flat potential 
for these fields. Following our previous work on the existence of scaling
solutions for the dilaton evolution in the heterotic string, 
which provided a solution to the so-call `runaway
problem' pointed out by Brustein and Steinhardt,
we investigated here the impact of a recent proposal to also
stabilize moduli of considering the
couplings of moduli to matter fields. After a thorough study of the issue
in the context of the heterotic string we concluded that these couplings
do not seem to affect stabilization for realistic string settings.

Next we considered M-theory scenarios, where there is also a runaway
problem associated, in this case, to the combination of dilaton and
modulus, $\Phi_-$, which has an exponential-type potential analogous
to that of the dilaton in the heterotic string. Here we study the
evolution equations for both $\Phi_+$ and $\Phi_-$ in the presence 
of a dominating background, in order to see whether we can also reach
a scaling regime that will make the fields settle at their minima
instead of rolling past them. It turns out that such a regime, i.e.
scaling, is achieved but at late stages of the moduli evolution, with 
the result that these fields are not stabilized while scaling, but in 
an intermediate regime. This makes it extremely difficult to perform
an analytic study of the problem but, nevertheless, we have been able
to determine a wide region of the parameter space (essentially defined 
by the initial positions of the fields, the type of 
background and the characteristics of the non-perturbative K\"ahler
potential) for which the stabilization is successful. Finally, we
considered the couplings of moduli to matter fields, and we concluded 
that the situation is the same as in the heterotic string case, i.e. 
these couplings do not contribute at all to improve the stabilization
problem of the moduli.

\vspace{1cm}

\section*{Acknowledgements}

We thank Ed Copeland for useful discussions. We would also like to 
thank Toni Riotto and the authors of Ref.~\cite{Huey00} for fruitful 
discussions on the issue of couplings to matter fields. The work of 
TB was supported by PPARC, and that of NJN by Funda\c c{\~a}o para
a Ci\^encia e a Tecnologia (Portugal). TB thanks the 
Theory Division at CERN for hospitality during the initial stages
of this work.


\end{document}